\documentclass[10pt]{article}
\usepackage{times}
\usepackage{graphicx}

\def\be{\begin{equation}}
\def\ee{\end{equation}}

\begin{document}

\begin{center}
{\Large\bf{
AN EXPERIMENTAL SETUP TO DEVELOP RFI MITIGATION TECHNIQUES
FOR RADIO ASTRONOMY}}
\vspace{3mm}

{\large\bf K. Jeeva Priya  and  D. Anish Roshi}
\vspace{3mm}

{\large\bf\itshape Raman Research Institute, Bangalore 560080 }
\vspace{3mm}
\end{center}

\noindent{\large\bf Abstract}
\vspace{3mm}

Increasing levels of Radio Frequency Interference (RFI) are a problem
for research in radio astronomy.  Various techniques to suppress
RFI and extract astronomical signals from data
affected by interference are being tried out. However, extracting
weak astronomical signals in the spectral region affected 
by RFI remains a technological
challenge. In this paper, we describe the construction
of an experimental setup at the Raman Research Institute (RRI),
Bangalore, India for research in RFI mitigation. We also present
some results of tests done on the data collected using this setup.
The experimental setup makes use of the 1.42 GHz receiver system
of the 10.4 m telescope at RRI. A new reference antenna, its receiver
system and a backend for recording digitized voltage together with
the 1.42 GHz receiver system form the experimental setup.
We present the results of the characterization of the experimental setup.
An off-line adaptive filter was successfully implemented and tested
using the data obtained with the experimental setup. \\

\noindent{\large\bf Introduction}
\vspace{3mm}

Radio frequency interference (RFI) is a growing problem for research
in radio astronomy. The ITU (International Telecommunication
Union) has defined and regulated the usage of the radio spectrum by
allocating various frequency bands for different
services including radio astronomy. Since radio
flux densities from cosmic sources are typically 40 to 100 dB below
those due to other services, often the out-of-band emission
from other services limits the sensitivity of astronomical
observations. Moreover, several radio spectral emissions
from cosmic sources are present in the frequency range
outside the allocated band for radio astronomy.
Thus mitigating interference and extracting
radio signals from cosmic sources is essential for advancement
of research in radio astronomy.

Several RFI mitigation techniques have been
developed and applied for radio astronomy application 
[1], [see also 2], [3 and references therein].
For example, a blanking algorithm to excise short (a few $\mu$sec wide) pulses
due to distance measuring equipment on astronomical data at frequencies
around 1 GHz was developed and applied to data taken using the Green Bank
telescope [4],[5]. A real-time adaptive cancellation technique to suppress
interference in the voltage domain was developed by [6], while [7]
developed a post-correlation technique, which essentially works on
the power domain (i.e. after detection). The deviation of the
probability distribution of the output of a
telescope from normal distribution in the presence of RFI was also used to
excise interference [8]. These techniques
are successful to a large extent, however, RFI rejection achieved
is not often sufficient for sensitive radio astronomy observations.
Thus research into developing new mitigation techniques is necessary. 
In this paper, we describe the construction of an experimental
setup at the Raman Research Institute (RRI) for developing
new mitigation techniques for radio astronomy application and present
some results of the test done on the data obtained with the setup. \\

\noindent{\large\bf The Experimental Setup}
\vspace{3mm}

\begin{figure}
\includegraphics[width = 6in,height = 3.5in,angle =0]{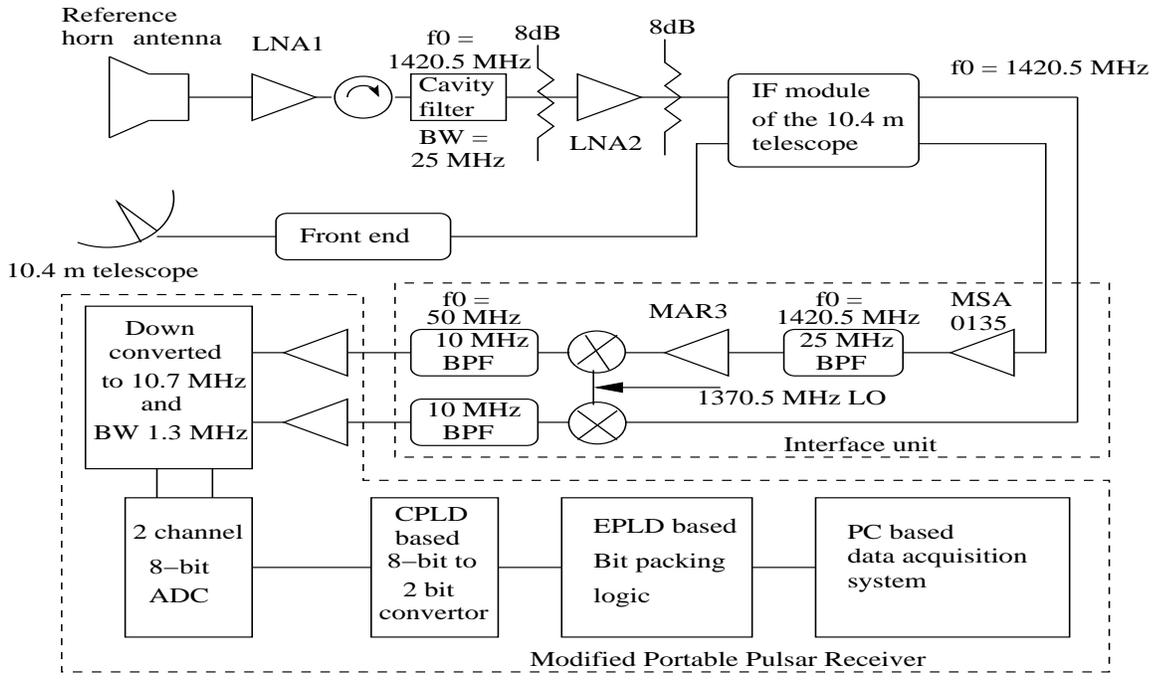}
\caption{A block diagram of the experimental setup.}
\label{fig1}
\end{figure}

A block diagram of the experimental setup is shown in Fig.~\ref{fig1}.
The 10.4 m telescope at RRI was originally built
for operation in the mm-wave band. A few years ago, a 1.4 GHz receiver
system was added to the telescope primarily for educational purposes.
The 1.4 GHz frequency band of this telescope is contaminated
with RFI and hence is an ideal system to develop and test interference
mitigation algorithms. We used one of the polarizations
of this receiver system as the `primary channel' to
receive the RFI contaminated astronomical signal.
We constructed a `reference channel' and used
the second polarization of the 1.4 GHz receiver system to
connect its output to the backend. The `reference channel' consists
of a pyramidal horn antenna, coupled to a low noise
amplifier (LNA) using a single linearized probe.
The horn antenna and the LNA operate near 1420.5 MHz.
The signal from the horn antenna is band limited to 25 MHz using a
cavity filter.
Both primary and reference channel outputs are further amplified
and down converted to 50 MHz in the interface unit. At this stage the
bandwidth is limited to 10 MHz. We modified an existing Portable
Pulsar receiver (PPR; [9]) to record the digitized voltage
from the two channels to a Personal Computer (PC) hard disk. The PPR
further limits
the bandwidth of the signals to 1.3 MHz. The
signals are bandpass sampled and digitized to 2 bits
before recording.

\begin{figure}
\includegraphics[width = 6in,height=3.5in,angle=0]{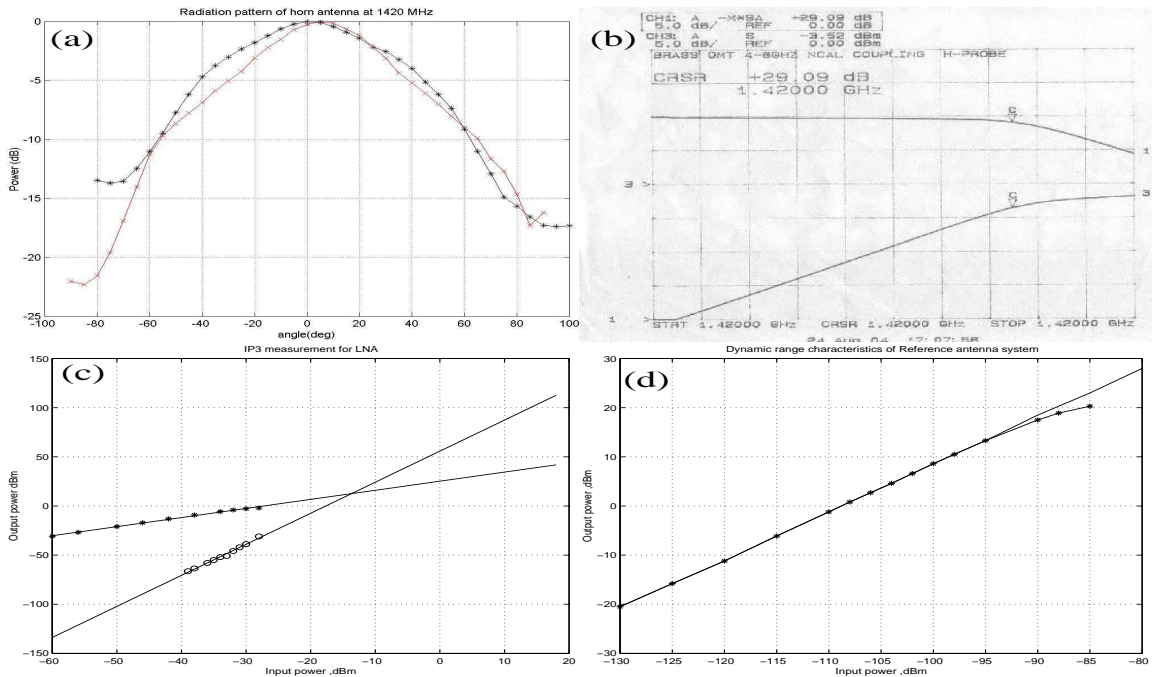}
\caption{Characteristics of the components of the experimental setup.
(a) E (black) and H (red) plane radiation patterns of the pyramidal horn
antenna. (b) Gain (top solid line marked 1) and output power (bottom solid
line marked 3) against input power to the LNA. The marker `c' indicates
29 dB gain, which corresponds to the 1 dB compression point.
The 1 dB compression point of the LNA is $-$3.5 dBm (marker `c' in the bottom
curve). (c) Result of the two-tone, third order intercept point measurement
of the LNA. The output IP3 of the LNA is +12.5 dBm. (d) Result of the
measurement of the 1 dB compression point of the analog part of the
reference receiver system. The measured 1 dB compression point is
+17 dBm. }
\label{fig2}
\end{figure}

The characteristics of the components of the experimental setup
were measured. The measured E and H plane radiation patterns of the pyramidal
horn antenna are shown in Fig.~\ref{fig2}. The half power beam
widths of the antenna in the E and H planes are 60$^o$ and 50$^o$
respectively. The LNA used in the `reference channel' is a
three stage HEMT amplifier [10], which has a measured gain of 30 dB,
noise temperature of 30 K and bandwidth of 500 MHz centered at 1.42 GHz.
The 1 dB compression point and the output IP3 (two-tone, third order
intercept point) of the LNA measured at 1.42 GHz are -3.5 dBm and
+12.5 dBm. We also measured the 1 dB compression point of the analog part
of the reference receiver system, which is +17 dBm.  \\

\noindent{\large\bf Preliminary Results}
\vspace{3mm}

\begin{figure}
\includegraphics[width = 6in,height=2.5in,angle=0]{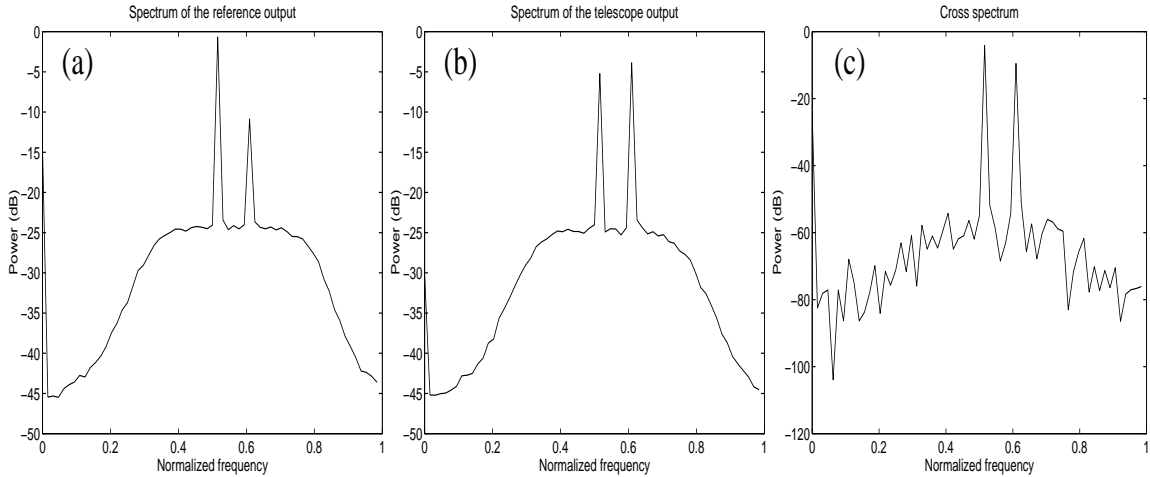}
\caption{Spectra obtained from the data recorded using the
experimental setup. Spectra obtained from the reference and telescope
outputs are shown in (a) and (b) respectively. Two narrow band
interference can be seen in the plots. The Fourier transform
of the cross correlation of the two signals is shown in (c).
All spectra are integrated for about 60 msec.}
\label{fig3}
\end{figure}
\begin{figure}
\includegraphics[width = 4in,height=1.5in,angle=0]{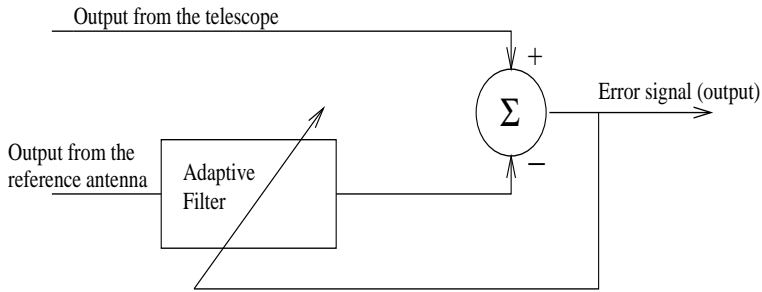}
\caption{A block diagram of the Matlab implementation of
the off-line adaptive filter. The 2 bit digitized data from
the experimental setup forms the inputs to the adaptive filter.
Fig.~\ref{fig5} shows the spectrum of the output of this filter.}
\label{fig4}
\end{figure}
\begin{figure}
\includegraphics[width = 6in,height=2.5in,angle=0]{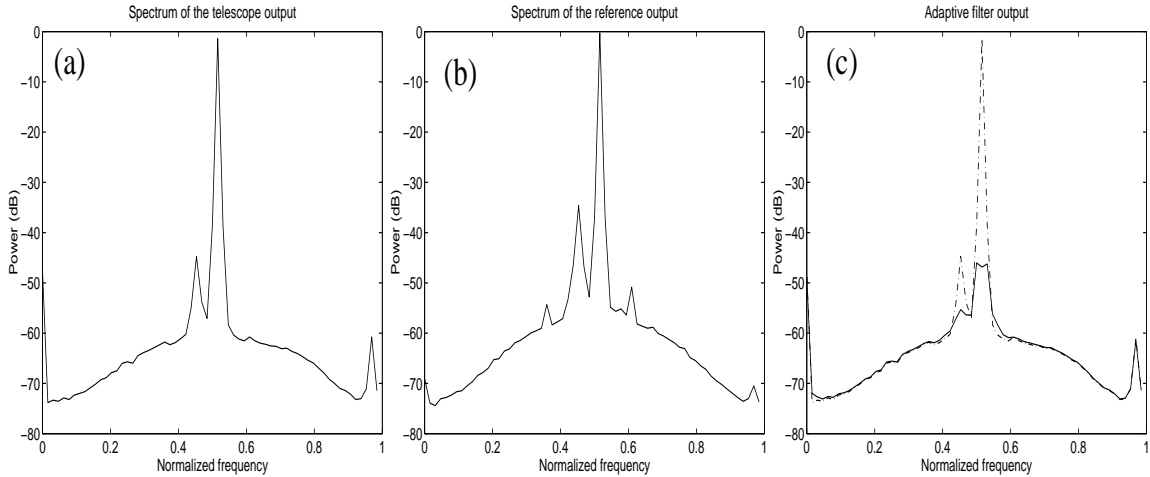}
\caption{Spectra of the telescope and reference antenna
outputs obtained with the radiated tone are shown in (a)
and (b) respectively. The spectrum of the output of the
adaptive filter is shown in (c). For comparison
the spectrum of the telescope output
is shown in `dash-dot' line in (c). As seen in the
figure the interference is attenuated after passing
through the adaptive filter. The radiated signal
(near normalized frequency of 0.5) is attenuated by
about 45 dB. }
\label{fig5}
\end{figure}

We  recorded the data using the experimental system by pointing the
telescope at different directions. Sample spectra obtained from
the recorded data are shown in Fig.~\ref{fig3}. The signals are sampled
at a frequency slightly larger than the Nyquist rate (ie 1.3 $\times$ 2 MHz) for
the purpose of producing Fig.~\ref{fig3}. Two strong narrow band interferences
near 0.5 and 0.6 normalized frequencies are present at the
telescope and reference antenna outputs. The cross correlation of
the two outputs confirms the common origin of both these interferences.

For testing the functioning of the experimental setup, we radiated
a single tone and recorded the data from both the reference and
telescope output. The recorded signals were passed through
an adaptive filter configuration similar to that developed
by [6]. The block diagram
of the adaptive filter, which was implemented using Matlab,
is shown in Fig.~\ref{fig4}. A 60 tap FIR
filter along with least mean square (LMS) algorithm was
used for the implementation. The spectra of the input signals  
to the adaptive filter and the spectrum of its output are
shown in Fig.~\ref{fig5}. The spectra were integrated for
about 100 msec. The narrow band signal near the normalized frequency
0.5 is the radiated tone. The narrow band signal near normalized
frequency 0.45 is an RFI from unknown source. As seen in Fig.~\ref{fig5},
both the radiated tone and the RFI are attenuated by the adaptive
filter. The RFI rejection obtained for the radiated tone is about 45 dB.
Note also that the spectral shape away from the narrow band
interferences is not
affected by the filtering process. \\

\noindent{\large\bf Acknowledgment}
\vspace{3mm}

We thank the staff of the radio astronomy laboratory and
mechanical workshop at RRI for their generous help during
the construction of the experimental setup. We
thank Prof. N. Udaya Shankar and Prof. C. R. Subrahmanya for
their support. The project described in this paper
is in partial fulfillment of KJP's M.E (Master of Engineering)
degree requirement. KJP thanks RRI for providing a project
assistantship. \\

\noindent{\large\bf Reference}
\vspace{3mm}

\noindent
[1]W. A. Baan, P. A. Fridman, R. P. Millenaar, ``Radio Frequency
Interference Mitigation at the Westerbork Synthesis Radio Telescope:
Algorithms, Test Observations, and System Implementation'', The
Astronomical Journal, Volume 128, pp. 933-949, 2004.

\noindent
[2]R. J. Fisher, ``RFI and How to Deal with It'', NAIC-NRAO School on
Single-Dish Radio Astronomy: Techniques and Applications, ASP conf.
series, eds. S. Stanimirovic, D. R. Altschuler, P. F. Goldsmith,
C. Salter, Vol. 278, p. 433-445, 2002.

\noindent
[3]P. A. Fridman, W. A. Baan, ``RFI mitigation methods in radio astronomy'',
Astronomy and Astrophysics, v.378, p.327-344, 2001.

\noindent
[4]J. R. Fisher, Q. Zhang, Y. Zheng, S. G. Wilson, R. F. Bradley,
``Mitigation of Pulsed Interference to Redshifted H I and OH
Observations between 960 and 1215 Megahertz'',
The Astronomical Journal, Vol. 129, pp. 2940-2949, 2005.

\noindent
[5]Q. Zhang, Y. Zheng, S. G. Wilson, J. R. Fisher, R. F. Bradley,
``Excision of Distance Measuring Equipment Interference from Radio
Astronomy Signals'',
The Astronomical Journal, Vol. 129, pp. 2933-2939, 2005.

\noindent
[6]C. Barnbaum, F. R. Bradley, ``A New Approach to Interference
Excision in Radio Astronomy: Real-Time Adaptive Cancellation'', The
Astronomical Journal, Volume 116, pp. 2598-2614, 1998.

\noindent
[7]F. H. Briggs, J. F. Bell, M. J. Kesteven,i ``Removing Radio
Interference from Contaminated Astronomical Spectra Using an
Independent Reference Signal and Closure Relations''
The Astronomical Journal, Volume 120, pp. 3351-3361, 2000.

\noindent
[8]P. A. Fridman, ``RFI excision using a higher order statistics
analysis of the power spectrum'', Astronomy and Astrophysics, vol.368,
pp. 369-376, 2001.

\noindent
[9]A. A. Deshpande et al., ``Portable Pulsar Receiver'', unpublished

\noindent
[10]A. Raghunathan, ``Building of the 21cm front end receiver for the
giant metrewave radio telescope'', M.Sc. Engg. Thesis, Bangalore
University, India, 2000

\end{document}